\documentclass[onecolumn]{emulateapj}
\usepackage{amsmath}
\usepackage{graphicx}
\usepackage{subfigure}
\usepackage{rotate}

\def\ltsima{$\;\buildrel < \over \sim \;$}
\def\simlt{\lower.5ex \hbox{\ltsima}}
\def\gtsima{$\;\buildrel > \over \sim \;$}
\def\simgt{\lower.5ex \hbox{\gtsima}}
\begin{document}
\title{Observations of Water Vapor Outflow from NML Cygnus}
\author{Viktor Zubko} 
\affil{NASA Goddard Space Flight Center code 913, Greenbelt, MD. 20771; also Science
Systems and Applications, Inc.}
\author{Di Li}
\affil{Harvard-Smithsonian Center for Astrophysics, 60 Garden Street, Cambridge, MA 02138}
\author{Tanya Lim } 
\affil{Rutherford Appleton Laboratory, Chilton, Didcot, Oxon, OX11 0QX}
\author{Helmut Feuchtgruber} 
\affil{Max-Planck-Institut f\"{u}r Extraterrestrische Physik, Postfach 1603, D-85740, Garching,
Germany}
\author{Martin Harwit}
\affil{511 H st., SW, Washington, DC 20024-2725; also Cornell University}

\begin{abstract}

We report new observations of the far infrared and submillimeter water vapor emission of NML
Cygnus based on data gathered with the Infrared Space Observatory and the Submillimeter
Wave Astronomy Satellite.  We compare the emission from NML Cyg to that previously
published for  VY CMa and W Hya in an attempt to establish the validity of recently proposed
models for the outflow from evolved stars.  The data obtained support the contention by
Ivezi\'{c} \& Elitzur (1995, 1997) that the atmospheres of evolved stars obey a set of scaling
laws in which the optical depth of the outflow is the single most significant scaling
parameter, affecting both the radiative transfer and the dynamics of the outflow.  Specifically, we
provide observations comparing the water vapor emission from NML Cyg, VY CMa and W Hya,
and find, to the extent permitted by the quality of our data, that the results are in reasonable
agreement with a model developed by Zubko \& Elitzur (2000).  Using this model we derive a
mass loss based on the dust opacities, spectral line fluxes, and outflow velocities of water vapor
observed in the atmospheres of these oxygen-rich giants.  For VY CMa and NML Cyg we also
obtain an estimate of the stellar mass.
\end{abstract}


\section{Introduction}

NML Cyg, also designated IRC+10448, has been studied extensively since its discovery four
decades ago in the near infrared sky survey conducted by Neugebauer, Martz \& Leighton
(1965).  Though the star is quite faint at visual wavelengths, it is enormously luminous in the
infrared.   Morris and Jura (1983) associated this star with the Cyg OB2 association, which
placed the star at $\sim 2$ kpc and defined its luminosity as $5\times 10^5$ $L_{\odot}$.  These
authors concluded that the star has a mass of $50$ $M_{\odot}$ and a minimum mass loss of
$\sim
6 \times 10^{-5}$ $M_{\odot}$ yr$^{-1}$, indicated by the dense dust shell surrounding the
star.  
The star exhibits significant variability in the infrared. Monnier et al (1997) report flux variations
of $\pm 0.25$ mag in the N band at 10.2 $\mu$m with an irregular period around 940 days. 

Using speckle interferometry in the near infrared in the K, L, and M bands, Ridgway et al.
(1986) estimated the optical depth at 3.4$\mu$m to be 2.4 and derived an inner radius of the dust
shell
of $\sim 0.045$ arcsec, where they estimated the dust temperature to be $\geq 1200$ K.  Through
an extensive de-reddening analysis these authors also found consistency with the previously
proposed spectral type M6 III (Wing, Spinrad \& Kuhi, 1967) and concluded that the star's 
effective temperature is $T_{eff}\sim 3250$\,K.  More recently, Monnier et al. (1997) conducted
long baseline interferometry at 11$\mu$m and modeled the dust outflow in somewhat greater
detail, obtaining better fits to all gathered data by assuming a stellar effective temperature of only
$T_{eff} \sim 2500$\,K.  

The star's line of sight velocity relative to the local standard of rest is approximately known. 
Cohen, et al (1987) investigated the 1612 MHz OH-maser emission from NML Cyg.  They
found the distribution among the many individual maser features they were able to resolve at
0.06 km s$^{-1}$ to be double peaked with peaks centered roughly at a velocity with respect
to the local standard of rest $v_{lsr} = \pm 22$ km s$^{-1}$,
suggesting that the central star's line of sight velocity is close to zero.   These authors assumed 
the star to lie at a distance of 2 kpc, which placed the radius of the maser emitting shell around
$7.3\times 10^{16}$ cm from the star.  The star's radius, based on its color temperature and
luminosity at that distance, was then $\sim 250$ times smaller, $\sim 3\times 10^{14}$ cm.
Diamond, Norris and Booth (1984), who extended these OH observations, confirmed a similar
stellar velocity.

A similar study concentrating on the water vapor masers associated with outflow from NML
Cyg by Richards et al. (1996) shows a rather different picture, exhibiting a strong component at
$v_{lsr} = -20$\,km s$^{-1}$, with no other emission besides a hint of a second feature at $\sim
+ 6$\,km s$^{-1}$.  This asymmetry is explained by the authors as due to a collimated outflow
close to the star at an angular distance of $\sim 100$ mas.  This outflow, they conjecture, 
gradually becomes less collimated at a larger distance of order a few arcseconds, where the OH
1612 MHz masers are observed. Observations carried out at several epochs indicate proper
motion of order 20 km s$^{-1}$, assuming a distance of 2 kpc.  This would be consistent both
with the radial velocity of the OH maser peaks and the approaching velocity of the H$_2$O
maser emission.

A paper by Danchi et al. (2001) reports on the dust distribution interferometrically measured
at 11$\mu$m in the infrared.  They conclude that there are two distinct dust shells, both of which
have moved away from the star by approximately the same distance over a period of $\sim 6$ yr.
Their separation suggests that they were formed $\sim 65$ yr apart.  If the transverse outflow is
comparable to the radial velocities of the masers, these authors favor a distance to the star of
$\sim 1200$\, pc, rather than 2 kpc, preferred by Richards et al. (1996).

The dust emission measured by Danchi et al. (2001) peaks at roughly $100$\,mas from the
star but is distributed rather broadly over distances from $\sim 25$ to $\geq 150$mas. 
Determining a displacement of only $\sim 25$\,mas of the diffuse dust outflow, measured over
the span of only 6 yr, appears to us to be a more difficult observation than measuring the angular
displacement of the more compact H$_2$O masers over a period of $\sim 13$\,yr, with quite
small error bars.  For purposes of our current paper, we therefore assume the distance of 2 kpc for
NML Cyg, suggested by Richards et al. (1996).

Boboltz \& Marvel (2000) have examined the $ v = 1, J = 1 \rightarrow 0$ SiO maser emission
around the star
with the Very Long Baseline Array.  They find that the masers form a roughly elliptical ring
around the star, with dimensions  $\sim 8\times 10^{14}$ by  $10^{15}$cm.  The maser
velocity distribution together with the spatial distribution led these authors to conclude that the
SiO maser shell rotates with a velocity $v\sin i \sim 11$ km s$^{-1}$, about a systemic
velocity of $-6.6$ km s$^{-1}$.  While it is difficult to see how such a high rotational velocity
could be induced at such a large distance from the star, a comparable system velocity emerges
from consideration only of the double-peaked distribution of maser velocities, with peaks at
$\sim 5$ and $-18$ km s$^{-1}$, averaging out to $\sim -6.5$ km s$^{-1}$.

The star's velocity can also be derived from its thermal SiO $v=0$ emission.  This has a roughly
parabolic spectral form, suggesting spherically symmetric outflow.   Lucas et al, (1992) found
that the peak emission occurs at about -5 km s$^{-1}$, and the mean emission velocity is  about
-1 km s$^{-1}$.  These authors also determined the radius at the half power point for SiO
emission to lie at $\sim 2\times 10^{16}$ cm, well beyond
the SiO maser emitting region but also well inside the shell of OH masers. They derived a
mass loss of order $1.6\times 10^{-4}$ $M_{\odot}$ for the star, somewhat higher, but not at
significant variance with the estimates of Morris and Jura (1983).

Many of these properties of NML Cyg are summarized in Table 1, where comparative values
for two other oxygen-rich stars, VY Canis Majoris and W Hydrae, can also be found.

\begin{table}[ht]
\begin{center}
\scriptsize
\begin{tabular}{l|c|c|c}
\hline \multicolumn{4}{c}{Table 1. Characteristics of VY CMa, W Hydrae \& NML Cyg}\\
\hline

Star   &  VY CMa $^a$& W Hydrae$^b$ & NML Cyg\\
Star's Distance (pc) & 1500 & 115 & 2000$^b$\\
Star's Luminosity ($L_{\odot}$) & $5\times 10^5$ & 11,050 & $5\times 10^5$\\
Star's Temperature (K) & 2800 & 2500 & 2500\ $^d$ \\
Star's Radius (cm) & $2.25\times 10^{14}$ & $3.9\times 10^{13}$ & $2.6\times 10^{14}\ ^d$\\
Final Outflow Velocity (km s$^{-1}$) & 20 & 10 & $\sim 25\ ^c$\\
Optical Depth & $A_J \sim 3.2^e$ & 0.83 at 5500\AA & $\sim 2$ at $11\mu$m$^b$\\
Gas-to-dust ratio & $\sim 100$ & 850 & $\sim 100\ ^b$\\
Mass loss rate $M_{\odot}$ yr$^{-1}$ & $\sim 2\times 10^{-4}$ & $2.3\times 10^{-6}$ &
$\sim 2\times 10^{-4}\ ^b$\\
\hline
\multicolumn{3}{l} {$^a$ Data from Neufeld et al. (1996) and Barlow et al. (1996)}\\
\multicolumn{3}{l} {$^b$ For sources of data see text}\\
\multicolumn{3}{l} {$^c$ NML Cyg data, this paper}\\
\multicolumn{3}{l} {$^d$ Model of Monnier et al. (1997)}\\
\multicolumn{3}{l} {$^e$ optical depth in the {\it J} band}\\
\hline
\end{tabular}
\end{center}
\normalsize
\end{table}

Since NML Cyg is one of the most luminous sources in the northern sky it was extensively used
as a calibration source for the Short Wavelength Spectrometer (SWS) Fabry-Perot (F-P) mode,
during the Infrared Space Observatory (ISO) mission.  Some of the spectral data reported here
were obtained as part of these calibration procedures.   On ISO the star was also studied as part of
several dedicated programs aimed at understanding the atmospheres of evolved oxygen-rich
stars. One of these programs concentrated on detecting the water vapor emission from such stars
with the aid of the Submillimeter Wave Astronomy Satellite (SWAS) as well as with SWS and
the Long Wavelength Spectrometer (LWS) on ISO.  ISO data were, respectively, obtained for W
Hya
by Barlow et al. (1996) and Neufeld et al. (1996), and for VY CMa by Neufeld et al. (1996,
1999), while SWAS data for these two stars were published by Harwit \& Bergin (2002).  Taken
together, these data tended to support the theoretical models for the atmospheres and outflow
from oxygen rich stars developed by Ivezi\'{c} and Elitzur (1995, 1997) and Zubko \& Elitzur
(2000).

Section 2 of the present paper reports water vapor observations obtained with ISO.  Section
3 reports similar observations obtained with SWAS. Section 4 briefly describes the theoretical
models for
stellar outflow devised by Zubko \& Elitzur (2000). Section 5 makes use of their model to obtain
a mass-loss rate for NML Cyg.  Section 6 compares the observed water-vapor line strengths to
those that the model predicts.  Section 7 discusses some apparent discrepancies in the
data and attempts to plausibly account for them.  In section 8 we discuss the balance
between gravitational attraction and repulsion by light pressure, which may account for some of
the apparent anomalies in outflow velocities and mass-loss rates. 

\section{ H$_2$O Observations with ISO}

Early water vapor observations of NML Cyg were obtained by Justtanont et al.
(1996) who used the SWS in grating mode and, in addition to other spectral features, detected
rovibrational transitions of  H$_2$O at 2.7 and 6.2 $\mu$m and 9 pure rotational lines of the
molecule between 36 to 45 $\mu$m.  The resolving power of the grating was of order 680 at
30$\mu$m --- insufficient to clearly determine line shapes.  From continuum observations and
an assumed outflow velocity of 27.7 km s$^{-1}$ these authors estimated a dust mass loss of
$\sim 2\times 10^{-6}$ $M_{\odot}$ yr$^{-1}$, which translates into an overall mass loss of 
$\sim 2\times 10^{-4}$ $M_{\odot}$ yr$^{-1}$ for a dust-to-gas mass ratio of 100.

The ISO observations cited in the present paper were carried out with the telescope pointed at
20$^{h}$46$^m$25.46$^s$, +40$^{\circ}$ 6' 59.6" (J 2000).
With the SWS/F-P, water vapor features were obtained at a spectral resolving power
of $\sim 30,000$.  Most of these mid-infrared lines emanate from high-lying energy levels of
H$_2$O and consistently exhibit P Cygni profiles, clearly seen at the higher spectral resolution
of the Fabry-Perot instrument.  The line shapes are consistent with the assumption that the lines
are
emitted close to the star where the stellar outflow is still at a high temperature, and that cooler
H$_2$O, moving at considerably higher outflow velocities at greater distances from the star,
absorbs
much of the emitted light.  Figure 1 shows several of the spectral lines registered with the
SWS/F-P, and shows the line fits that we adopted.

At a resolving power of 30,000 the SWS/F-P emission lines are well resolved.  In order to
recover the emission line strength before self-absorption by cool water vapor in the outer parts of
the outflow, we assumed that the source spectrum consists of a continuum flux $F_c$ plus a
parabolic emission feature $F = F_0\cdot( v-v_0)^2/v_e^2$ for the range $|v-v_0| < v_e$, where
$v_0$ is the line of sight velocity of the star and $v_e$ is the expansion velocity in the emitting
region.  This velocity is expected to be different for different emission lines, and provides
information on the expansion velocity in different temperature regimes.

We then fitted the parabola convolved with the instrument profile to most closely replicate the
portion of the emission feature on the long-wavelength side of the peak.  Since this portion of the
emission line should be least affected by self-absorption in an accelerating outflow,  this
procedure permitted us to restore the emitted flux on the short wavelength side of the emission
peak and estimate the total emitted flux.  These steps tended to increase the line flux by 20 to
30\% relative to the directly measured emission component in the P Cygni profiles.
 To fully  simulate the P Cygni profile, we further
postulated a constant fractional absorption $f_{abs}$ over some velocity range $v_{min}$ to
$v_{max}$ applied to the continuum plus fully restored parabolic emission.  All three
parameters, $f_{abs}$, $v_{min}$ and $v_{max}$ were allowed to be free.   For a given value
of $f_{abs}$ we selected $v_{min}$ to give the best fit to the short-wavelength side of the
absorption feature and chose $v_{max}$ to give the best fit to the short-wavelength side of the
emission feature and the long-wavelength side of the absorption feature. Between these three
parameters we generally obtained a good fit.  The peaks of the emission lines tend to fall around
$v_{lsr}= -10$. We find $v_{max}$ to be of order $-50$ km s$^{-1}$ for most of our fits, while
the star's line of sight velocity with respect to the local standard of rest is taken to be
$v_{lsr}\sim 0$ km s$^{-1}$.  Such a high value for $v_{max}$ is somewhat surprising, as
discussed in Section 7, below.

The P Cygni profiles in the mid-infrared water vapor features of NML Cyg are consistent with
similar profiles observed in VY CMa (Neufeld et al., 1999), while the same  lines
observed by Neufeld et al. (1996) in W Hya show no self-absorption, indicating a significantly 
lower column depth in the star's cool outer layers.

The spectral lines observed by the LWS/F-P were obtained at a spectral resolving power of $\sim
10,000$.  These lines  involve lower-lying energy states emitted by cooler gas at greater distances
from the star, and exhibit no noticeable self absorption.  Presumably this is due to their emission
in the more tenuous outer regions of the outflow where self-absorption is less likely.   Figure 2,
shows four emission lines obtained with the LWS/F-P.  The data were reduced with the ISO
LWS/F-P interactive data reduction program. At 95 $\mu$m the resolving power of the
instrument is $\sim 9800$ and gradually declines to about $9500$ at 125 $\mu$m.  

To obtain these data we repeatedly stepped the LWS Fabry-Perot instrument through a
wavelength range that spanned roughly 7 to 8 steps on either side of the line which generally
encompassed another 4 to 6 steps.  These roughly 20 to 22 steps were equally spaced.   Each step
was displaced approximately half a spectral resolution element or $\sim 15$ km s$^{-1}$ from
the previous step.  The entire spectral range was stepped through repeatedly in what was
designated as the ``rapid-scanning mode" of operation.  In reducing the data, a sloping spectral
base line was subtracted.  The reduced spectra are shown in Figure 2. 
The best-fit spectral peak is displaced to shorter wavelengths relative to the laboratory values,
with a consistent velocity displacement of $12 - 14$ km s$^{-1}$, in good agreement with the
SWS/F-P values.  Full widths at half maximum range from $\sim 31$ to $\sim 69$ km s$^{-1}$. 
Because the spectral resolving power in this wavelength range, $\sim 30$\, km s$^{-1}$, is
comparable to the observed line widths, and  significantly broader than the spectral velocity
resolution of the SWS spectra or the SWAS spectra reported in Section 3, we fitted the observed
emission features with a single Gaussian shape plus a baseline, a choice primarily determined by
the Fabry-Perot transmission function.  This line fit exhibits broader wings than a parabolic fit
might suggest, but appears more appropriate under the circumstances and, in any case, does not
significantly affect the results we report.  

\section{H$_2$O Observations with SWAS}

NML Cyg was observed in the 557 GHz $1_{10} - 1_{01}$ {\bf ($538 \mu$m)} ground state
ortho transition
with SWAS.  The spectral resolution with this instrument is of order $3\times 10^5$, i.e. about 1
km s$^{-1}$.  The observed line is considerably weaker than the infrared lines, but nevertheless
was detectable.  The data were gathered in observations covering 165 hours during two
different observing sessions.  The first covered 106 hours spreading over the period from
November 8, 2002  to December 20, 2002.  The second observing session covered 59 hours and
took place
between April 26 and May 15, 2003.  The emission line together with a parabolic fit is
shown in Figure 3.  The spectral resolving power in this spectral range is more than an order
of magnitude higher than in the SWS and LWS ranges, and the observed line shape is dominated
entirely by the actual outflow.  The parabolic fit we have chosen reflects the expectation that the
outflow may, to first order,  be isotropic in the star's rest frame. The flux integrated over the line
is $1.3\times 10^{-20}$ W cm$^{-2}$, the peak flux is 184
Jy and the peak of the emission line lies at $v_{lsr} = 0$ km $s^{-1}$. The one sigma
uncertainty
of the fitted integrated flux is  $0.18\times 10^{-20}$ W cm$^{-2}$.

The emission observed in this spectral range is due to cold water vapor.  The upper level of
this transition lies only $T = h\nu/k\sim 26.7$\,K above the ground  ortho-H$_2$O state, so that
it is readily excited in collisions with cool gas in the outer regions of the star's outflow.

\section{Theoretical Models of the Outflow from Evolved Stars}

Until the past few years, models of the outflow from evolved stars generally treated the radiative
transfer through the outflowing atmosphere separately from the dynamics of the flow.  This
tended to lead to conflicting notions:   On the radiative transfer side, the outflow
was considered to proceed at some constant terminal velocity. From the dynamic perspective,
dust grains were pictured as absorbing starlight which they quickly re-emitted in the infrared
while also transferring absorbed momentum to the gas, thus accelerating the outflow.  But if the
infrared radiation is so closely tied to momentum transfer, it cannot simultaneously be made
consistent with a constant outflow velocity --- leading to the cited conflict.

This difficulty was first overcome in a series of papers by Ivezi\'{c} \& Elitzur
(1995, 1997) and  Zubko \& Elitzur (2000), who produced a self-consistent outflow model
combining both dust radiative transfer and hydrodynamics.   In oxygen-rich evolved stars, water
vapor is a dominant coolant in the outflow.  The model exhibits scaling laws that define the
radial variations in the velocity and the radiation fields in terms of a radial distance measured in
units equal to the inner diameter of the dust absorption shell.   The gas in
the outflow is largely heated through collisions with the grains that also transfer linear
momentum to the gas, thereby accelerating it.  Over a significant
portion of the outflow the balance between the heating and cooling is largely dominated by 
water vapor.  The H$_2$O is mainly heated by absorption of radiation emanating from dust in
the immediate vicinity of the star, though some heating also is due to collisions with hydrogen
molecules that have been collisionally heated by dust grains. H$_2$ cannot rapidly cool through
radiative transitions since
it is restricted to quadrupole emission at the temperatures found in outflows.  Most of the
heating and cooling takes place through pure rotational H$_2$O transitions throughout the bulk
of the outflow, the cooling being almost directly proportional to the heating throughout the flow
(Zubko \& Elitzur, private communication, 2001). As discussed by Harwit \& Bergin (2002) this
nearly constant proportionality has the effect of simulating a radiative transfer and temperature
profile of a gray body, scaling the temperature $T$ to radial distance $R$ from the star through
the constancy of the product $T^4R^2$.

\section{Derived Mass Loss from NML Cyg and VY CMa}

Figure 4 shows representative points from the infrared continuum spectrum of
NML Cyg obtained through ground-based photometry (Strecker \& Ney 1974),
with additional data from IRAS.  These points are fitted with a synthetic spectrum modeled on
the procedure of Zubko \& Elitzur (2000), who assumed a silicate-rich dust envelope in the
outflow from oxygen-rich stars and used the DUSTY  code of Ivezi\'{c}, Nenkova, \& Elitzur
(1999). Figure 5 provides matching data for VY CMa.  The observational data are obtained from
Harwit et al. (2001). Input parameters for the computations that produce the fitted spectra of
Figures 4 and 5 are shown for the respective stars in Table 2.

The spectral fits to the observed dust continuum spectra shown in Figures 4 and 5, while not
ideal, seem satisfactory, particularly given the high variability of each star.  The temperature and
outflow velocity profile derived with the model of Zubko \& Elitzur is shown for NML Cyg and
VY CMa in
Figure 6.  Again, it is worth noting the roughly constant value of $T_{{\rm gas}}^4 R^2$,
already mentioned in Section 4.

\begin{table}[ht]
\begin{center}
\scriptsize
\begin{tabular}{lcc}
\hline \multicolumn{3}{c}{Table 2.  Input, Fitting, and Derived  Parameters for NML Cyg and
VY CMa}\\
\hline
                                          & NML Cyg                     &  VY CMa\\
\hline\\
{\bf Input Parameters:}                   &                             &  \\
Stellar BB temperature $T_{\ast}$         &  2500 K                     &  2800 K\\
Luminosity $L_{\ast}$                     &  $5 \times 10^5 L_{\odot}$  &  $5 \times 10^5
L_{\odot}$\\
Distance                                  &  2000 pc                    &  1500 pc\\
Terminal outflow velocity$^{a, b}$ $v_e$  &  25 km s$^{-1}$             &  20 km s$^{-1}$\\
{\bf Fitting Parameters:}                 &                             &  \\
Optical depth $\tau_v$ at 0.55$\mu$m      &  75                         &  75\\
Dust temperature at $R_{\rm{inner}}$         &  1400 K                     &  1400 K\\
$Y \equiv R_{\rm{outer}}/R_{\rm{inner}}$       &  1000                      &  1000\\
H$_2$O abundance at $R_{\rm{inner}}$         &  $4 \times 10^{-5}$         &  $4 \times
10^{-4}$\\
Ortho-H$_2$O : para-H$_2$O                &  1                          &  1\\
{\bf Derived Parameters:}                 &                             &  \\
Mass loss rate $\dot M$                   &  $6.9 \times 10^{-4}$       &  $9.0 \times 10^{-4}$\\
Gas-to-dust mass ratio                    &  350                        &  510\\
Stellar radius                            &  $2.6 \times 10^{14}$ cm   &  $2.1 \times 10^{14}$ cm\\
Shell inner radius $R_{\rm{inner}}$          &  $1.05 \times 10^{15}$ cm   &  $1.15 \times
10^{15}$ cm\\
\hline
\multicolumn{3}{l}{$^a$ Silicate dust optical data from Laor \& Draine (1993)}\\
\multicolumn{3}{l}{$^b$ Dust size distribution from Mathis, Rumple \& Nordsieck (1977)}\\
\hline
\end{tabular}
\end{center}
\normalsize
\end{table}

The outflow models constructed here for NML Cyg and VY CMa are essentially different from
the one established by Zubko \& Elitzur (2000) for W Hydrae, in that the wind is optically thick
for both NML Cyg and VY CMa, with $\tau_v \sim 70$ to 80, whereas for W Hydrae, $\tau_v
\sim 1$.  This means that the dust envelope defines the total flux in most of the spectrum, as
indicated in Figures 4 and 5 by the merger, beyond $\sim 10 \mu$m, of the dotted line
representing dust emission and the solid line fitted to the total observed flux.  

Table 2 underlines the similarities between NML Cyg and VY CMa.  One major difference,
however, is
the ten times higher H$_2$O abundance in VY CMa, required by the star's large observed
H$_2$O line fluxes.  

To check the stability of our results for NML Cyg, we varied the temperature of the central star
over the range $2,500 \pm 400$K, the terminal outflow velocity over the range $25 \pm 10$km
s$^{-1}$, and the optical depth $\tau_v$ over values of $75 \pm 10$, constrained
largely by the near infrared portion of the spectrum in the 1 - 8 $\mu$m region.  Combining these
variations, with dust temperatures $1400\pm 100$ K at the inner edge of the outflow, well within
the realm constrained by the far infrared spectrum in the 20 to 100 $\mu$m region, and a ratio of
outer to inner edge of the outflow $Y = 1,000\pm 100$, we derived a range of outflow rates of
$\dot M = 7\pm 2\times 10^{-4}$ $M_{\odot}$\, yr$^{-1}$ and a gas-to-dust ratio $350\pm 30$.
Our estimates of $\dot M$ and final velocity, however, are subject to an additional
uncertainty of $\sim 30\%$, and the gas-to-dust ratio to an additional uncertainty of $\sim 60\%$,
due to the way that the DUSTY model takes gravity into account.  This is further discussed in
section 8.  Given the quality of available data, these results appear to be satisfactorily robust.
The uncertainties in the fitting and derived parameters for VY CMa are approximately the same.

Values cited in the ``Model" columns of Table 3 provide our best predictions of integrated line
fluxes, especially for spectral lines not yet observed.  Future observations thus will be able to
pass judgement on the general validity of the model.

\begin{table}[ht]
\begin{center}
\scriptsize
\begin{tabular}{c|c|c|c|c|c|c|c|c|c}
\hline \multicolumn{10}{c}{Table 3. Relative H$_2$O Flux Levels from W Hydrae, VY CMa
and NML Cyg}\\
\hline
Wavelength          & Transition & $T_u$ & $T_{\ell}$ & \multicolumn{2}{c}{W Hydrae}           
   & \multicolumn{2}{|c|}{VY CMa}                 & \multicolumn{2}{c}{NML Cyg}                \\
$\lambda$\ ($\mu$m) &            &  (K)  &    (K)     &
\multicolumn{2}{c}{$10^{-20}$W\,cm$^{-2}$} &
\multicolumn{2}{|c|}{$10^{-19}$W\,cm$^{-2}$} &
\multicolumn{2}{c}{$10^{-20}$W\,cm$^{-2}$} \\
\hline
                    &                               &      &      & Observed$^a$ & Model$^b$ & Observed$^c$ &
Model$^d$ & Observed & Model$^d$ \\
\hline
538.289             & o: $1_{10}\rightarrow 1_{01}$ & 61.0 & 34.2 & 0.45 & 0.76 & 0.34     &
0.57 & 1.3      & 1.2\\
179.527             & o: $2_{12}\rightarrow 1_{01}$ & 114  & 34.2 & 8.66 & 12.5 & 8.7      & 5.7 
&          & 16.5\\
174.624             & o: $3_{03}\rightarrow 2_{12}$ & 196  & 114  & 9.21 & 8.3  & 8.8      & 3.8 
&          & 10.7\\
125.356             & p: $4_{04}\rightarrow 3_{13}$ & 319  & 204  & 16.6 & 12.4 &          & 5.2 
& 14.6     & 14.1\\
113.538             & o: $4_{14}\rightarrow 3_{03}$ & 323  & 196  & 17.4 & 11.3 &          & 5.6 
& 10.1     & 14.7\\
108.073             & o: $2_{12}\rightarrow 1_{10}$ & 194  & 61.0 & 13.0 & 15.5 & 13.2     & 8.5 
& 17.5(??) & 19.1\\
100.983             & p: $2_{20}\rightarrow 1_{11}$ & 196  & 53.4 &      & 22.7 &          & 10.6 &
32.9     & 21.8\\
99.493              & o: $5_{05}\rightarrow 4_{14}$ & 468  & 323  &      & 12.6 &          & 5.6  &
40.0     & 15.3\\
95.627              & p: $5_{15}\rightarrow 4_{04}$ & 469  & 319  &      & 12.2 &          & 5.2  &
14.3     & 15.1\\
89.989              & p: $3_{22}\rightarrow 2_{11}$ & 297  & 137  & 27.3 & 18.1 &          & 8.8 
& 24.5(??) & 20.1\\
78.742              & o: $4_{23}\rightarrow 3_{12}$ & 432  & 249  & 28.0 & 14.1 &          & 7.4 
& 37.8(??) & 17.7\\
66.438              & o: $3_{30}\rightarrow 2_{21}$ & 411  & 194  & 22.9 & 18.5 & 17.2(??) &
10.3 &          & 21.6\\
58.699              & o: $4_{32}\rightarrow 3_{21}$ & 550  & 305  & 19.5 & 15.4 &          & 8.9 
& 52.5(??) & 18.8\\
40.760              & p: $6_{33}\rightarrow 5_{24}$ & 952  & 599  &      & 22.4 & 11.7(??) & 11.3
& 53       & 24.9\\
$^e$40.691          & o: $4_{32}\rightarrow 3_{03}$ & 550  & 196  & 23.0 & 36.1 & 36.9(??) &
20.2 &          & 43.6\\
37.566              & p: $7_{44}\rightarrow 6_{33}$ & 1335 & 952  &      & 13.7 &          & 8.3  &
37       & 15.8\\
36.212              & p: $6_{24}\rightarrow 5_{15}$ & 867  & 469  &      & 33.7 & 5.6(??)  & 17.2
& 48       & 36.7\\
35.938              & o: $6_{52}\rightarrow 5_{41}$ & 1279 & 878  &      & 13.9 & 12.1(??) & 9.0 
& 47       & 18.3\\
$^e$31.772          & o: $4_{41}\rightarrow 3_{12}$ & 702  & 249  & 63.0 & 40.0 & 24.1     &
24.6 &          & 47.5\\
29.837              & o: $7_{25}\rightarrow 6_{16}$ & 1126 & 643  & 32.0 & 30.7 & 23.5     &
19.8 &          & 39.6\\
\hline
\multicolumn{10}{l}{$^a$ ISO from Neufeld et al. (1996), Barlow et al. (1996);
SWAS data, this paper and Harwit \& Bergin (2002)}\\
\multicolumn{10}{l}{$^b$ W Hydrae model from Zubko \& Elitzur (private communication,
2001)}\\
\multicolumn{10}{l}{$^c$ For sources of data see text}\\
\multicolumn{10}{l}{$^d$ This paper}\\
\multicolumn{10}{l}{$^e$ Blending may need to be taken into account}\\
\multicolumn{10}{l}{(??) Question marks indicate less reliably established line fluxes}\\
\hline
\end{tabular}
\end{center}
\normalsize
\end{table}

\section{Comparison of Theory and Observations}

Once a best-fit outflow model was in hand, we again used the Zubko \& Elitzur (2000) model to
derive the expected line strengths for all the water vapor lines, based on a best fit to the relatively
few actually observed water vapor features, respectively, in NML Cyg, VY CMa and W
Hydrae.    

Because the model does not take into account the self absorption of emitted radiation in the
stellar outflow, the SWS emission lines exhibiting P Cygni profiles have to be adjusted, as
described in Section 2,  to recover the originally emitted flux level.  In Table 3 we present the
water vapor emission line data for NML Cyg, W Hya, and VY CMa obtained with ISO and
SWAS, along with the model reductions that Zubko \& Elitzur developed for W Hya (Harwit
\& Bergin, 2002) and similar new calculations for NML Cyg and VY CMa.  The model assumes
that scaling laws apply, and we do find rough self-consistency in that respect.  Table 3 shows that
the water vapor spectra of the three stars, despite their considerable differences in luminosity
and distance appear to
approximately scale relative to each other, VY CMa providing line fluxes that
are roughly ten times higher than W Hya and about four times higher than NML Cyg.

\section {{\bf Some Apparent Anomalies}}

Somewhat surprising in the data on NML Cyg is the rather large displacement of the absorption
trough seen in the P Cygni profiles of mid-infrared lines originating in regions of high
temperature surrounding the star.  As Figure 1 shows, some of the inferred outflow velocities in
the absorbing gas reach values of the order of 40 km s$^{-1}$, though with a mean value more
nearly 30 km s$^{-1}$.  These velocities are higher than the displacement of the far-infrared
lines observed in the LWS Fabry-Perot mode which uniformly are centered on roughly -15 km
s$^{-1}$ and have full widths at half maximum ranging from $\sim 30$ to 70 km s$^{-1}$ for
different, somewhat noisy line shapes. The SWAS data at 557 GHz also suggest lower outflow
velocities; the observed line, though weak, appears centered  on $v_{lsr} \sim 0$ with a full
width at half maximum around 40 km s$^{-1}$.  Comparable velocity profiles were observed by
Lucas et al. (1992) for the pure rotational $J = 2\rightarrow 1$ thermal SiO emission at 86.85
GHz.  A somewhat similar, though less marked discrepancy in line velocities also appears for
VY CMa, as judged from the data presented by Neufeld et al. (1999) and Harwit \& Bergin
(2002).  

For NML Cyg, the prime discrepancy in the outflow velocities lies between ISO data, which
concentrated on lines emitted in higher temperature regions, and the data on lower temperature
regions gathered by SWAS and in observations of SiO thermal emission.  These apparent
discrepancies could be resolved if some of the hot absorbing gas near the star were streaming out
at a substantially higher velocity than cooler gas at larger distances. Diamond, Norris and Booth
(1984) found similarly puzzling results in their observations of 1612 MHz OH maser emission
from NML Cyg.  They suggested the existence of an inner component  with an expansion
velocity of 34 km s$^{-1}$ with an extent of 1.5 arcsec and an outer component with an
expansion velocity of 20 km s$^{-1}$ at an angular radius of 2.5 arcsec.  Even the inner of these
two regions, however, would lie at a distance of $\sim 5\times 10^{16}$ cm from the star and, if
in local thermal equilibrium, would be far too cool to excite the levels responsible for the P
Cygni absorption.  Nevertheless, it is possible that different expansion components, even in the
inner regions surrounding the star, are flowing out at different velocities.  This could also explain
the substantial differences in line widths found for the various LWS emission lines. The larger
widths could be  an artifact of the line-fitting procedure, since the data are quite noisy. However,
they might also be real, reflecting velocity differences at different radial distances from the star. 
In future searches, it would be useful to see whether substantial velocity differences exist at
different radial distances in the outflow from NML Cyg.

Our mass loss rate for NML Cyg and VY CMa is also substantially higher than that cited by
other observers.  While Morris and Jura (1983) only give a minimum mass loss of $6\times
10^{-5}M_{\odot}$ yr$^{-1}$ for NML Cyg, this is an order of magnitude lower than the value
we give in Table 2.  Their calculation is based on the idea of an inverse Str\"{o}mgren sphere,
i.e. the ionization of the outer edge of the outflow from NML Cyg by ultraviolet radiation from
Cyg OB2 star 5.  The authors assume this ionizing star to be at the same distance from
Earth as NML Cyg, i.e. at the minimum possible distance from NML Cyg,  $\sim 100$ pc, given the two stars' angular separation of $\sim 2.9^{\circ}$.  The radial distance of the ionization front
from NML Cyg is then assumed to lie on a circumstellar surface at which the ionization rate just
balances the outflow rate.  The mass loss rate computed in this fashion can only be a minimum,
and scales as the square of the actual distance between the stars.

Knapp et al. (1982) used observations of the J = 1 -- 0 line of CO, by Zuckerman et al.
(1977), to determined a mass loss of $1.8\times 10^{-4}M_{\odot}$ yr$^{-1}$ for an outflow
velocity of 21 km s$^{-1}$, if their assumed stellar distance is changed from 200 pc to 2 kpc. 
Their mass loss rate is based on the assumption that the observed CO flux is produced at a
surface where heating by collisions with ambient gas keeps the CO in thermal equilibrium with
that gas.  Beyond this radial distance and the angular disk that it subtends on the sky, the flux is
assumed to drop to zero.  The calculated mass loss rate then becomes independent of the
[CO]/[H$_2$] ratio.

A third estimate of the mass loss for NML Cyg, due to Netzer \& Knapp (1987), is $\sim
1.6\times 10^{-4}M_{\odot}$ yr$^{-1}$, when adjusted to a distance of 2 kpc.  It is based on the
production of OH in the photodissociation of H$_2$O by the interstellar radiation field at the
outer surface of the outflow from the star.  The observed OH maser shell radii together with the
wind outflow velocity then provide a mass loss rate, on the assumption that the abundance of
water vapor in the outflow is [H$_2$O]/H$_2$] $= 3\times 10^{-4}$. The calculated mass loss
also depends on the assumed interstellar radiation field. The estimated error bars the authors
place on their mass loss rates are plus or minus a factor of 2.  However, the assumption of a
typical interstellar radiation field, appears to be in conflict with the ionizing radiation field
assumed by Morris \& Jura (1983).  It is not clear which of the two assumptions, if either, is
correct. 

\section{Discussion}

The mass loss of the three stars considered here is generally assumed to be largely driven by
the starlight absorbed by dust grains.  The approach we have taken in this paper is to calculate the
mass loss rate derived from the coupled dynamic and radiative transfer equations, based largely
on dust opacity (Ivezi\'{c}, Nenkova, \& Elitzur, 1999).  The assumptions we make are thus quite
different from those of Moris \& Jura (1983), Knapp et al. (1982), or Netzer \& Knapp (1987). 
The luminosity L of a star, when totally absorbed by dust, can lead to an absorption of outward
directed momentum amounting to L/c per second.  If the outflow accelerates from a negligibly
low initial velocity to high final velocity, v, as indicated in our Figure 6, then the amount of mass
that can be accelerated to a velocity v in each second is $\dot M = L/cv$.  Taking the assumed 
luminosity $L = 2\times 10^{39}$ erg/sec and a final velocity of 25 km/sec, leads to $\dot M =
2.67 \times 10^{22}$\,g s$^{-1}$ or $4\times  10^{-4} M_{\odot}$ yr$^{-1}$.  This is of the
order the mass loss indicated by the computer model we used.  

Our model assumes a spherical outflow and that all of the star's light is absorbed by the dust. 
That the outflow is spherical seems to be in accord with the parabolic line shape that Harwit \&
Bergin (2002) found for the 557 GHz outflow from VY CMa, and the generally parabolic line
fits for stars with massive outflows observed for many other stars, e.g. in the CO J = 2 -- 1 line
(Knapp et al., 1982).  About the nearly total absorption of star light, there also can be little doubt,
judging from the optical depths observed for NML Cyg and VY CMa. Two factors, however, do
enter consideration.  The first is the initial outflow velocity of gas just before dust condensation
and the onset of radiation driven outflow.  The second is the gravitational attraction of the star.

The peak outflow velocity $v_{inner}$ of the gas beyond the inner radius where dust is formed is not currently known,
but it appears to be high.  The spectral lines emitted and absorbed in the hottest portions of the
outflow, i.e. nearest to the star have considerably higher velocities than the flow at larger
distances.  This can be explained if gravitational attraction plays a decisive role once a packet of
gas has been accelerated to peak velocity and then is shielded from further radiative acceleration
by a new dust layer arising between it and the star.

While the DUSTY computer model of Ivezi\'{c}, Nenkova, \& Elitzur (1999) appears to be
the most advanced program available at the moment, an inherent limitation is that it takes the
force of gravitational attraction into account as a reduction in the radiative repulsive force
through a parameter $\Gamma = {\cal F}_{rad}/{\cal F}_{grav}$. $\Gamma$ is then taken to be
constant at all distances.  This is a valid approximation for optically thin outflows. But when the
optical depth of an outflow is high, the radiation pressure at large distances from the star drops
drastically as dust at large distances from the star no longer is able to absorb the far-infrared flux
reradiated near the inner radius of the outflow.  In this case the gravitational attraction of the star
can become dominant beyond the inner edge of the dusty outflow where the radiation pressure
drops. We, therefore, may expect a reduction in outflow velocities at large distances from the
star, a property not reflected in our computer generated outflow velocity in Figure 6.

If we take the short wavelength outflow velocities $v_{inner}$ to be equal to $(L/(c\dot M) +
v_0$, where $v_0$ is the outflow velocity at $R_{inner}$, the inner edge of the dust cloud where
the dust forms, then the final outflow velocity $v_f$, at large distances from the star, determined
by the half-width of the 557\,GHz line, will be determined by 
\begin{equation}
v_f = v_0 + \frac {L}{c\dot M} - \left(\frac{2MG}{R_{inner}}\right)^{1/2}\ .
\end{equation}
For NML Cyg, we can set $v_f\sim 19$\,km s$^{-1}$, and $v_0 + L/(c\dot M) = v_{inner} =
42$\,km s$^{-1}$.  This tells us that $2MG/R_{inner}\sim 23$\,km s$^{-1}$.  Choosing
$R_{inner} \sim 10^{15}$ cm then yields the mass of the star as $M\sim 20 M_{\odot}$. Cooling of
the gas will not change this outcome significantly, since the internal energy of gas at $T\sim
1400$\,K, characteristic of the temperature at $R_{inner}$, is only of order
$6\times 10^{10}$\,erg g$^{-1}$.  We can also solve for the mass loss, as 
\begin{equation}
\dot M \sim \frac{L}{c(v_{inner} - v_0)}
\end{equation}
For NML Cyg, we then have $\dot M\gtrsim 2.4\times 10^{-4} M_{\odot}$ yr$^{-1}$, on the
assumption that $v_0$ must be greater than or equal to zero if the gas is to recede sufficiently far
from the star to permit dust to form. 

Corresponding velocities for VY CMa are $v_f\sim 17$\,km
s$^{-1}$ and $v_{inner}\sim 30$\,km s$^{-1}$.  As in Table 2 we also take $R_{inner}\sim 1.15\times
10^{15}$ cm.  These parameters lead to $M\sim 8 M_{\odot}$ and $\dot M\gtrsim 3.3\times 10^4
M_{\odot}$\,yr$^{-1}$.  

For both stars we expect $v_0$ to be significantly lower than $v_{inner}$, implying that $v_{inner}$ is dominated by radiative acceleration. This is supported by the findings of Richards et al. (1996) who noted significant acceleration in the H$_2$O maser outflows of NML Cyg, suggesting a low value of $v_0$.  If so, the lower limit to the mass outflow is close to the actual value.

It, therefore, appears that both the total mass loss and the diminishing velocities as a function
of distance from the star can be roughly accounted for. Within factors of $\sim 2$, the mass loss
rates we derive appear consistent with those of other workers, who derived their values from
widely differing assumptions.  Our mass estimate for VY CMa seems low, unless this star has
already lost a substantial fraction of its initial mass.  Unfortunately, we still are some way from
narrowing these
differences and reaching a satisfactory understanding of massive outflows.

\section*{Acknowledgments}

This work was supported by NASA contract NAS5-30702. We would also like to
acknowledge our indebtedness to the referee for the careful reading of our manuscript and
incisive comments.

\vfill\eject
\centerline{\bf References}
\vskip 0.1 true in 
{\hoffset 20pt
\parindent = -20pt

Barlow, M. J., et al. 1996, A\&A 315, L241

Boboltz, D. A., \& Marvel, K. B. 2000, ApJ, 545, L149
                                                       
Cohen, r. J., et al., 1987, MNRAS, 225, 491

Danchi, W. C., et al. 2001, ApJ 555, 405

Diamond, P. J., Norris, R. P., \& Booth, R. S. 1984, MNRAS, 207, 611

Harwit, M., et al. 2001, ApJ, 557, 844

Harwit, M., \& Bergin, E. A. 2002, ApJ, 565, L105

Ivezi\'{c}, \v{Z}., \& Elitzur, M. 1995, ApJ, 445, 415

Ivezi\'{c}, \v{Z}., \& Elitzur, M. 1997, MNRAS, 287, 799

Ivezi\'{c}, \v{Z}., Nenkova, M., \& Elitzur, M. 1999, User Manual for
DUSTY, Univ. Kentucky Internal Report (Accessible at
http://www.pa.uky.edu/~moshe/dusty)

Justtanont, K., et al. 1996, A\&A, 315, L217

Knapp, G. R., et al. 1982, ApJ 252, 616

Laor, A. \& Draine, B. T. 1993, ApJ 402, 441

Lucas, R., et al. 1992, A\&A, 262, 491

Mathis, J. S., Rumple, W., \& Nordsieck, K. H. 1977, ApJ, 217, 425

Monnier, J. D., et al. 1997, ApJ, 481, 420

Morris, M., \& Jura, M. 1983, ApJ, 267, 179

Netzer, N. \& Knapp, G. R. 1987, ApJ 323, 179

Neufeld, D. A., et al. 1996, A\&A, 315, L237

Neufeld, D. A., Feuchtgruber, H. Harwit, M., \& Melnick, G. J. 1999, ApJ, 517 L147 

Neugebauer, G., Martz, D.E., \& Leighton, R.B. 1965, ApJ, 142, 399

Richards, A. M. S, Yates, J. A., \& Cohen, R. J. 1996, MNRAS 282, 665

Ridgway, S. T., et al. 1986, ApJ 302, 662

Strecker, D. W., \& Ney, E. P. 1974, \aj, 79, 1410

Wing, R. F., Spinrad, H., \& Kuhi, L.V. 1967, ApJ, 147, 117 

Zubko, V., \& Elitzur, M., 2000, ApJ, 544, L137

Zuckerman, B., et al. 1977, ApJ 211, L97

}

\begin{figure}[htp]
     \centering 
     \subfigure{\includegraphics[width=8cm, angle=0]{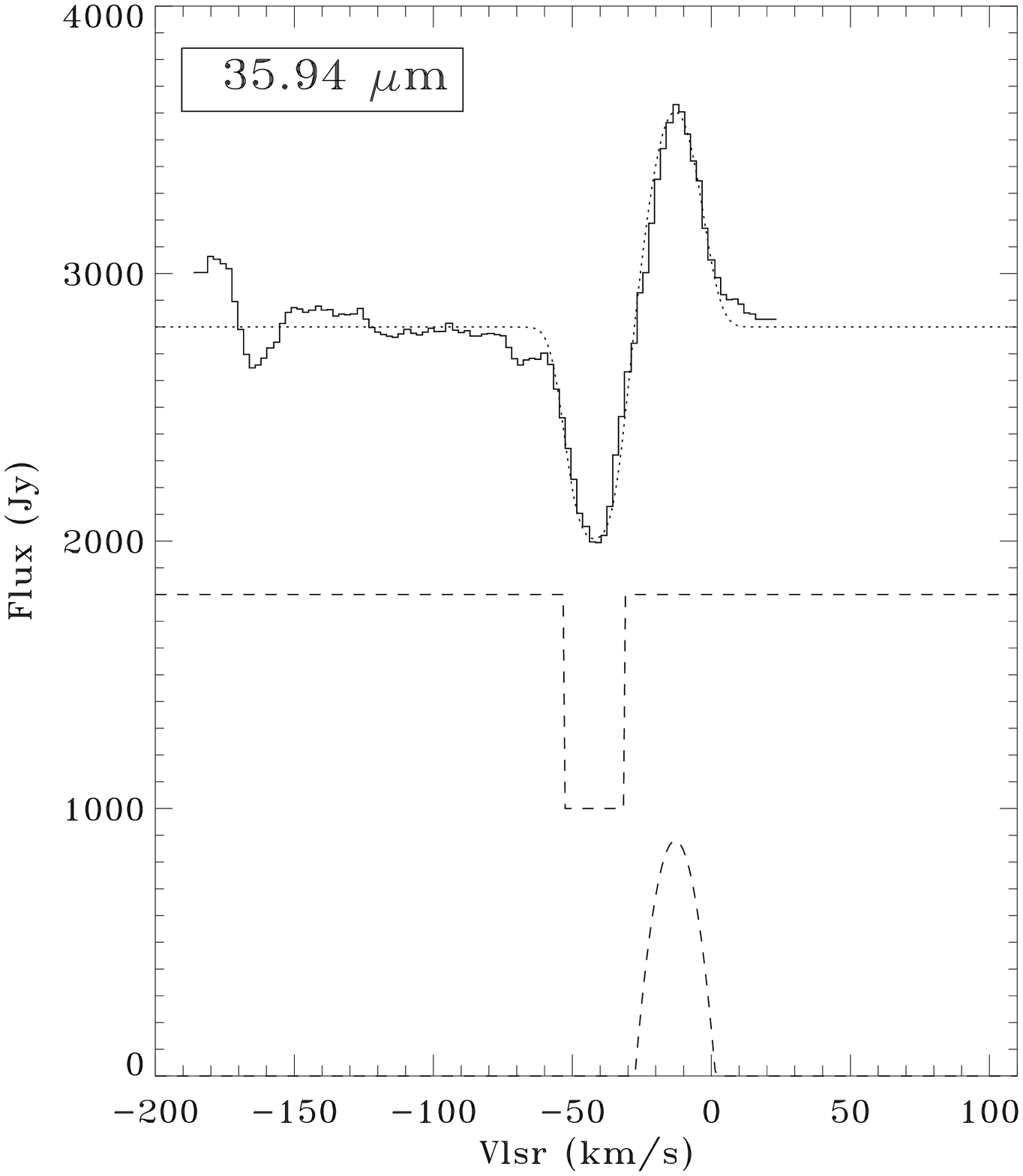}}
     \subfigure{\includegraphics[width=8cm, angle=0]{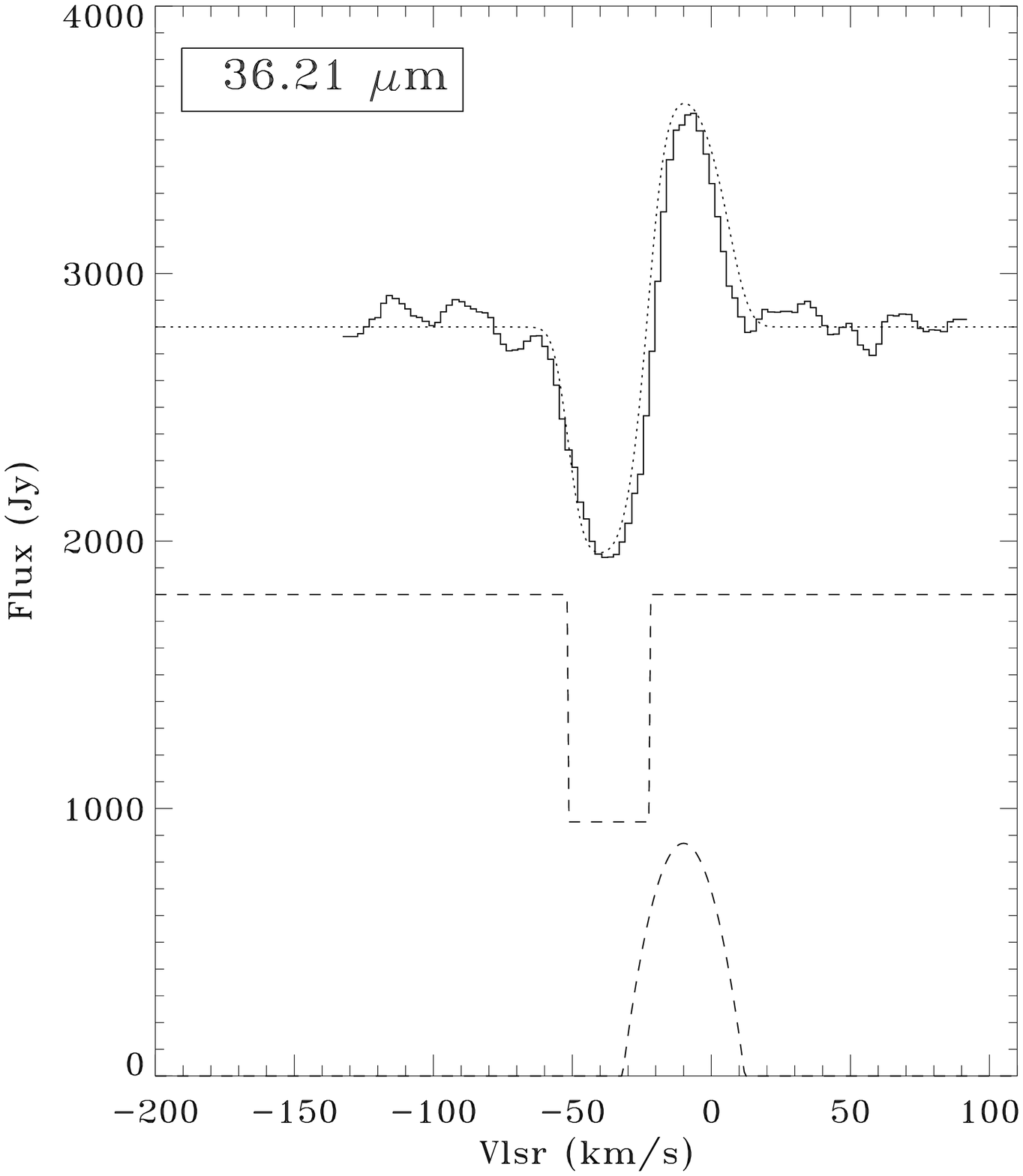}}\\
     \subfigure{\includegraphics[width=8cm,angle=0]{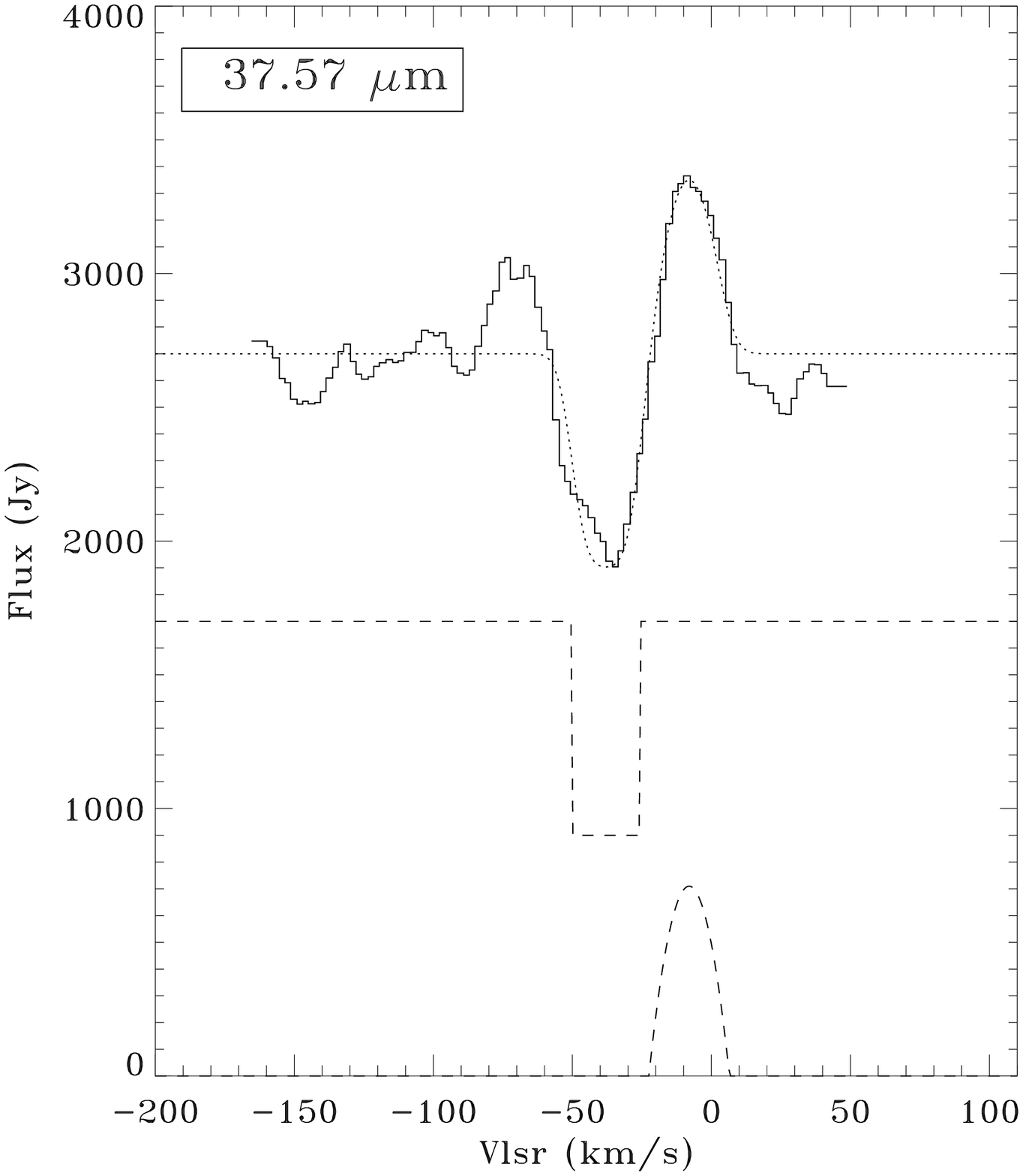}}
     \subfigure{\includegraphics[width=8cm, angle=0]{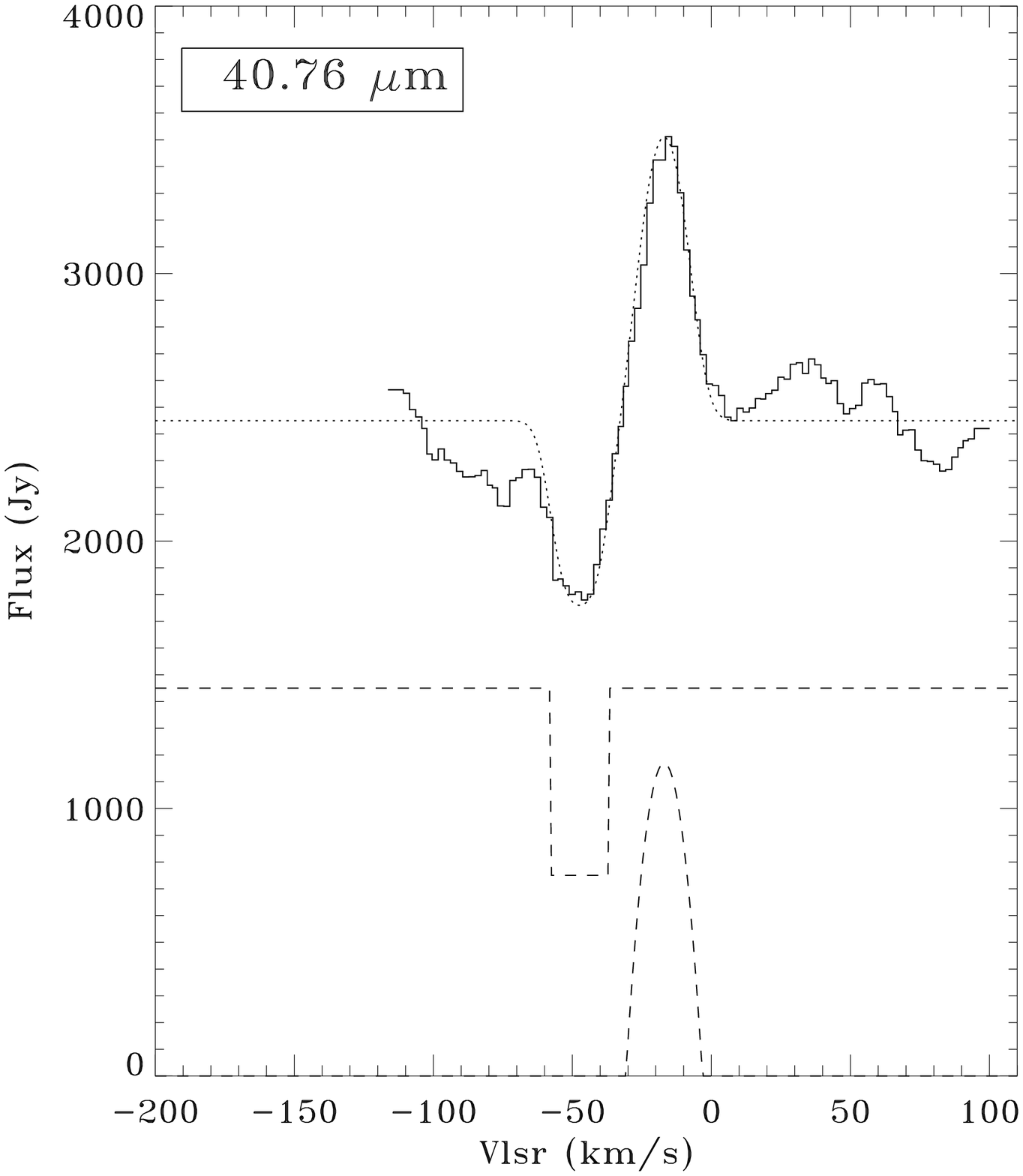}}
\caption{The spectra and model fits for data from the ISO SWS/F-P instrument.
In all panels the wavelength of the H$_2$O transition is indicated in the
upper left corner. The solid dark lines are ISO data from the SWS/F-P
instrument. The dashed lines represent the two components of our model, the 
absorbed continuum (displaced downward by 1000 Jy in order to be seen more clearly)
and the optically thick emission from an expanding shell 
(see text for more details). The dotted lines show the models convolved with instrument
response.}
\label{fig:SWS}
\end{figure}


\begin{figure}[htp]
     \centering
  \subfigure{\includegraphics[width=7cm,angle=0]{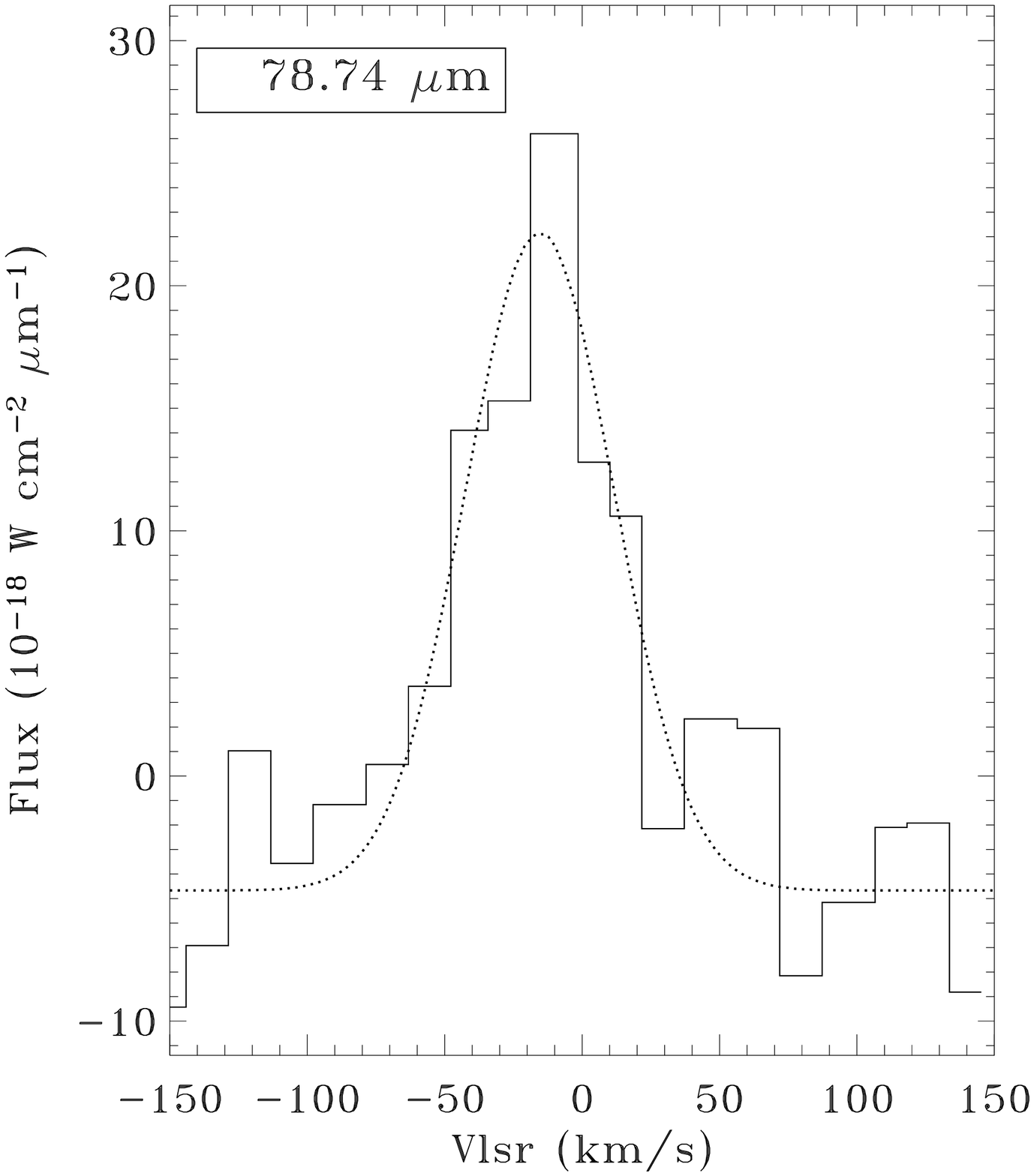}}
     \subfigure{\includegraphics[width=7cm, angle=0]{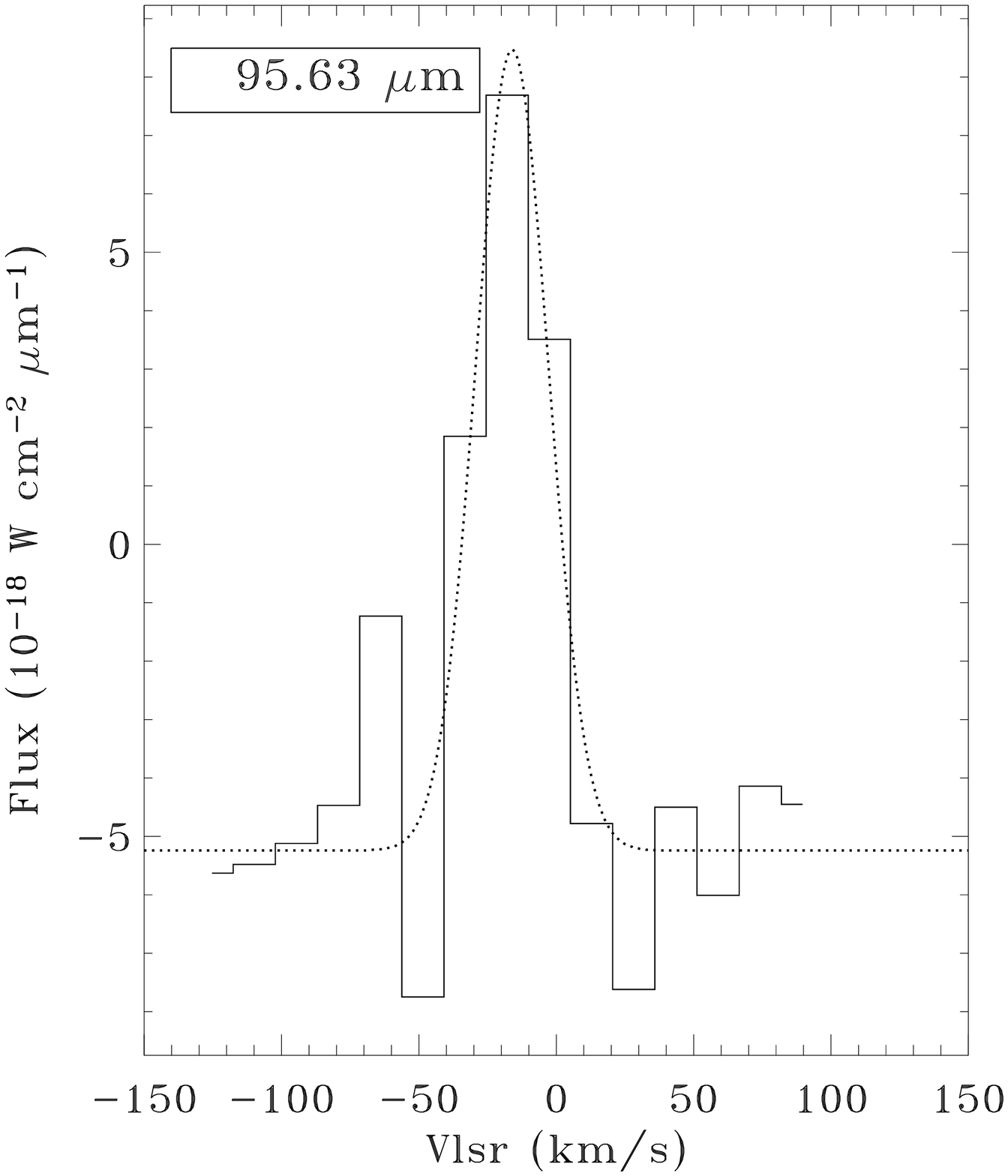}}\\
     \subfigure{\includegraphics[width=7cm, angle=0]{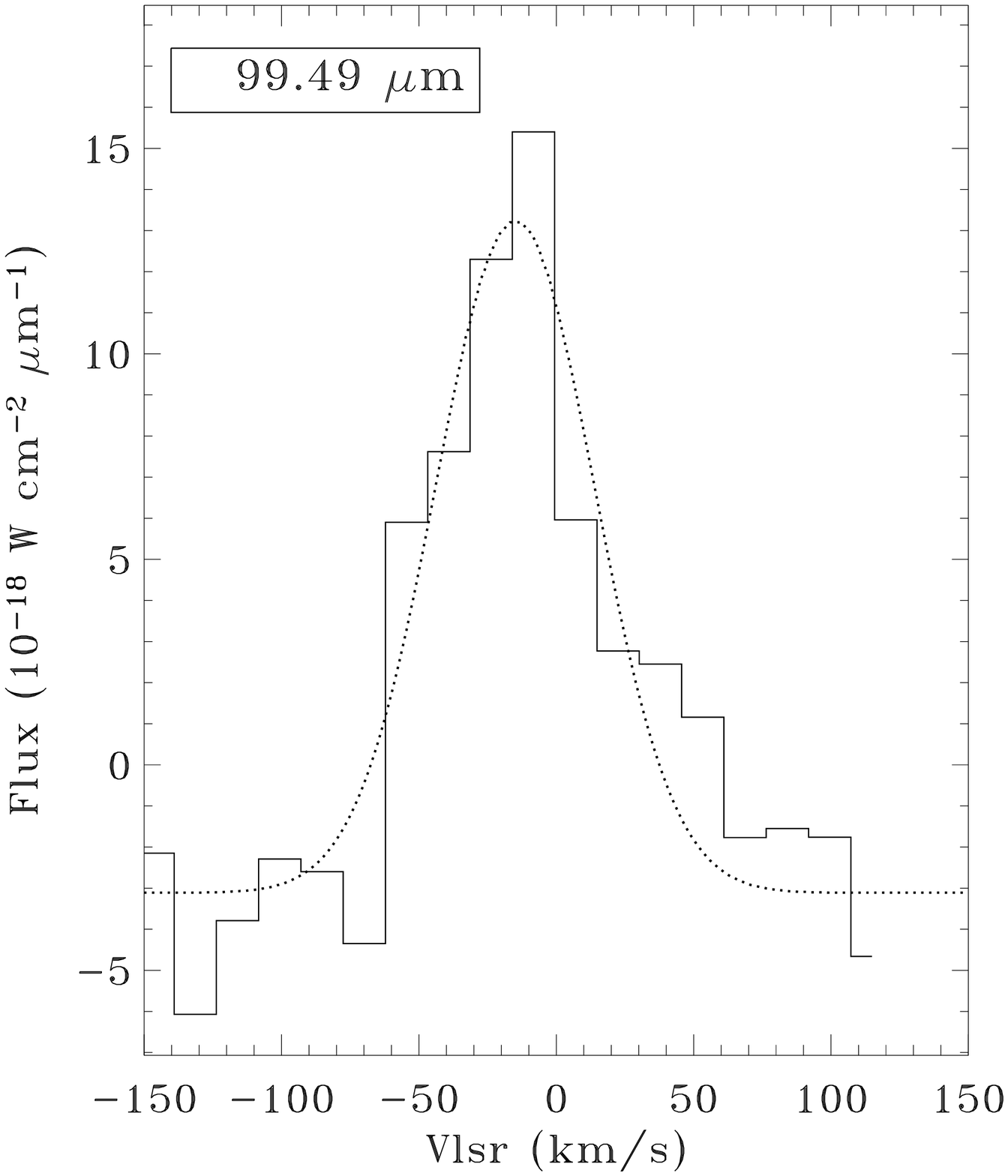}}
     \subfigure{\includegraphics[width=7cm, angle=0]{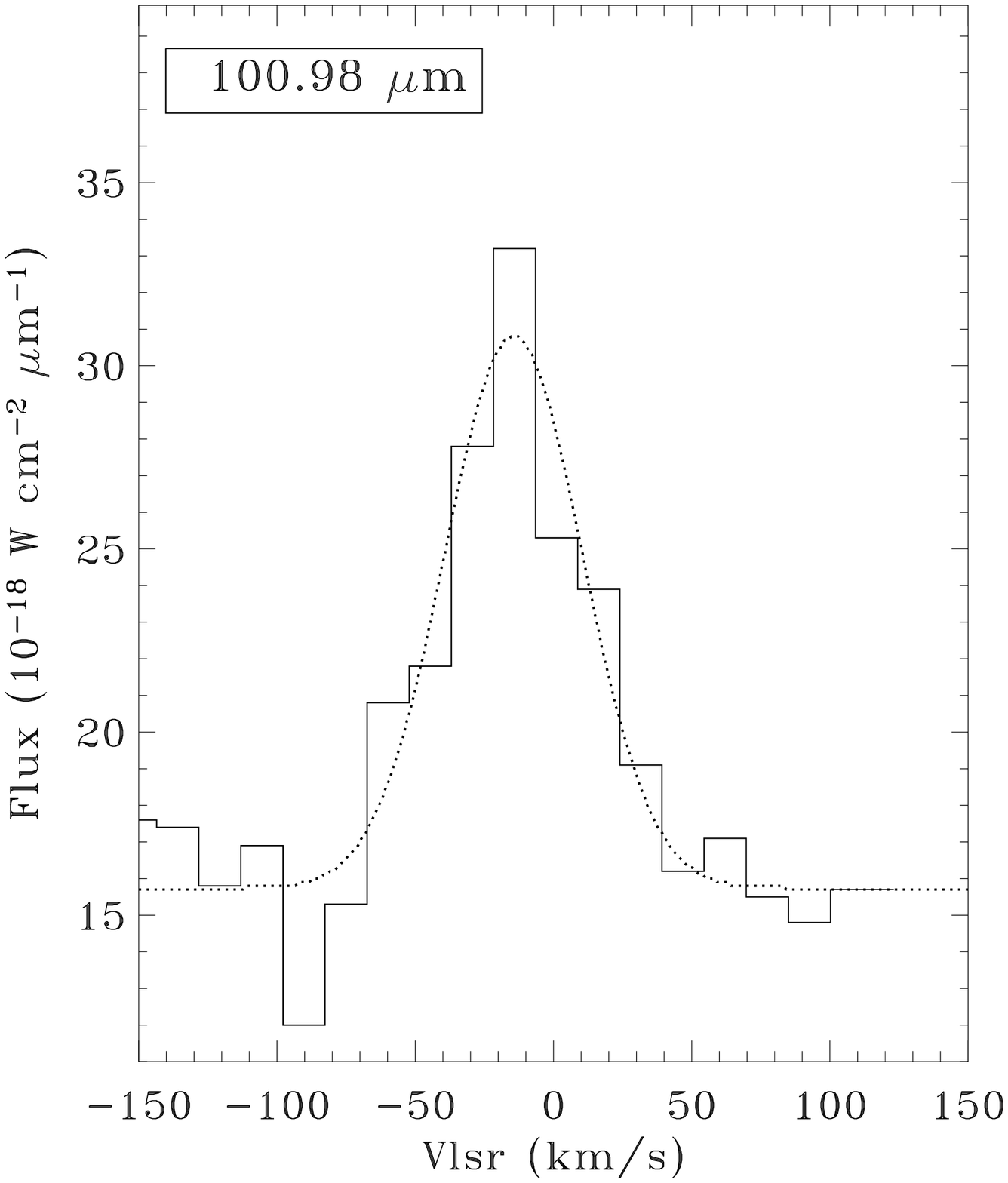}}\\
 \caption{The spectra and model fits for data from ISO LWS. The laboratory wavelengths of the
spectral lines are indicated in the box on each plot. The dotted lines show the fit with a single
Gaussian shape plus a baseline.}
\label{fig:LWS}
\end{figure}

\begin{figure}[htp]
     \centering 
     \subfigure{\includegraphics[width=12cm, angle=-90]{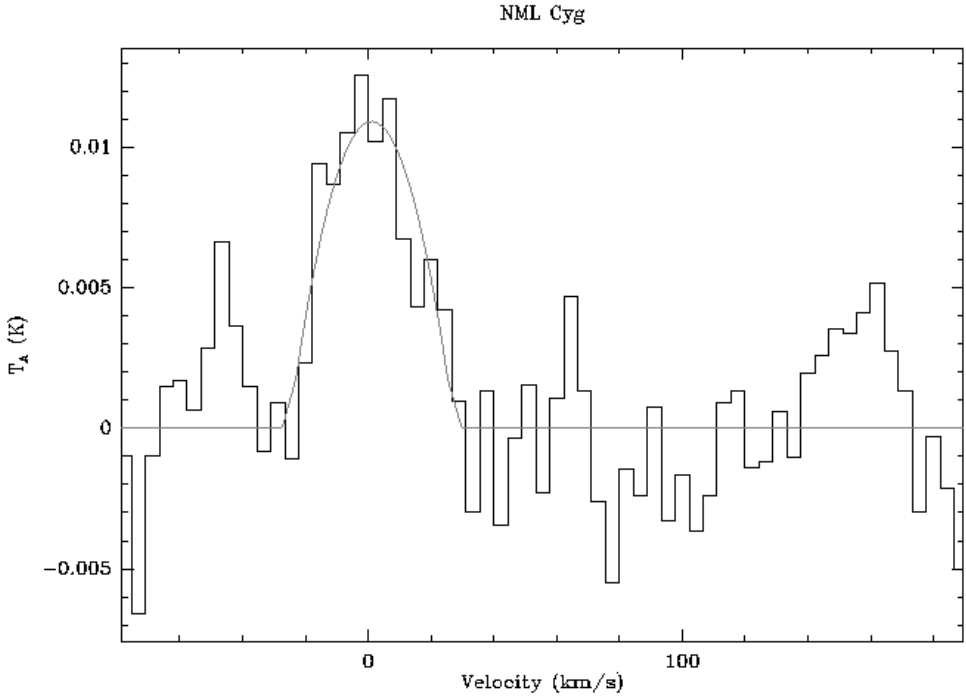}}
\caption{The SWAS data of the ortho H$_2$O $1_{10}\rightarrow 1_{01}$
transition at 538.289 $\mu$m, smoothed to a channel width of 4.5 km/s. The dotted line
shows a fit of an optically thick shell with an expansion velocity of 26 km/s. The displayed
velocity is with respect to the local standard of rest.}
\label{fig:SWAS}
\end{figure}

\begin{figure}[htp]
                  \centering 
     \subfigure{\includegraphics[width=12cm, angle=0]{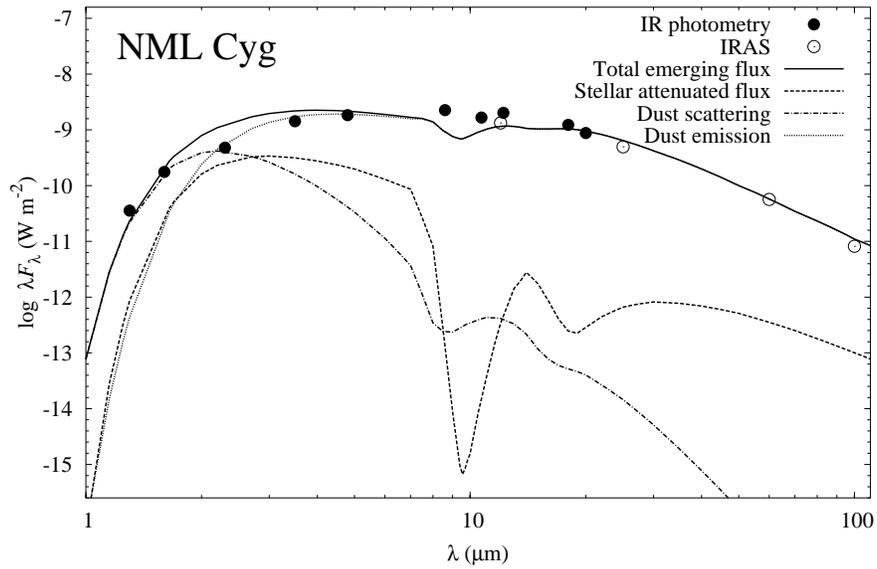}}
\caption{The continuum spectrum of NML Cyg.  Modeled components are identified in the
legend. Photometric data indicated by filled circles are from Strecker \& Nye (1974). }
\label{fig:Outflow}
\end{figure}

\begin{figure}[htp]
     \centering 
     \subfigure{\includegraphics[width=12cm, angle=0]{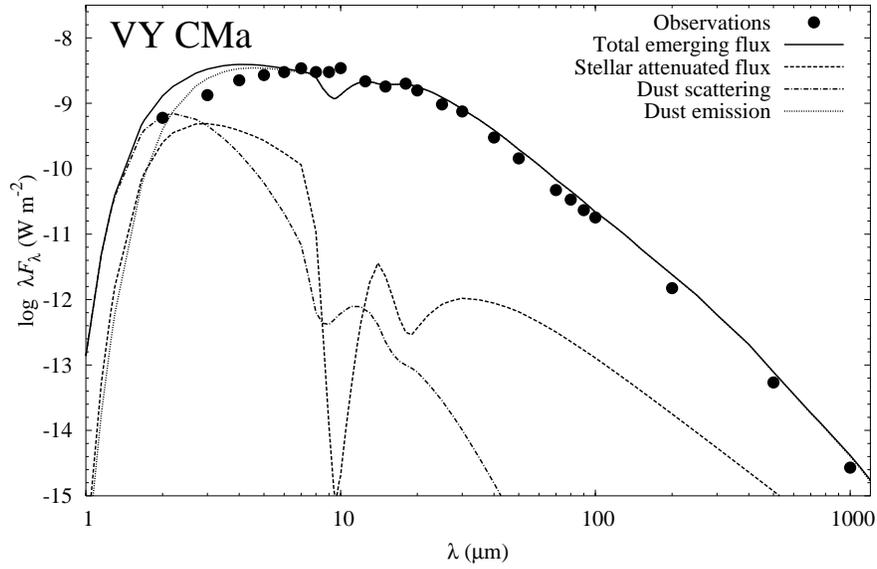}}
\caption{The continuum spectrum of VY CMa.  Modeled components are identified in the
legend. Photometric data indicated by filled circles are from Harwit et al. (2001). }
\label{fig:Outflow}
\end{figure}

\begin{figure}[htp]
     \centering 
     \subfigure{\includegraphics[width=18cm, angle=0]{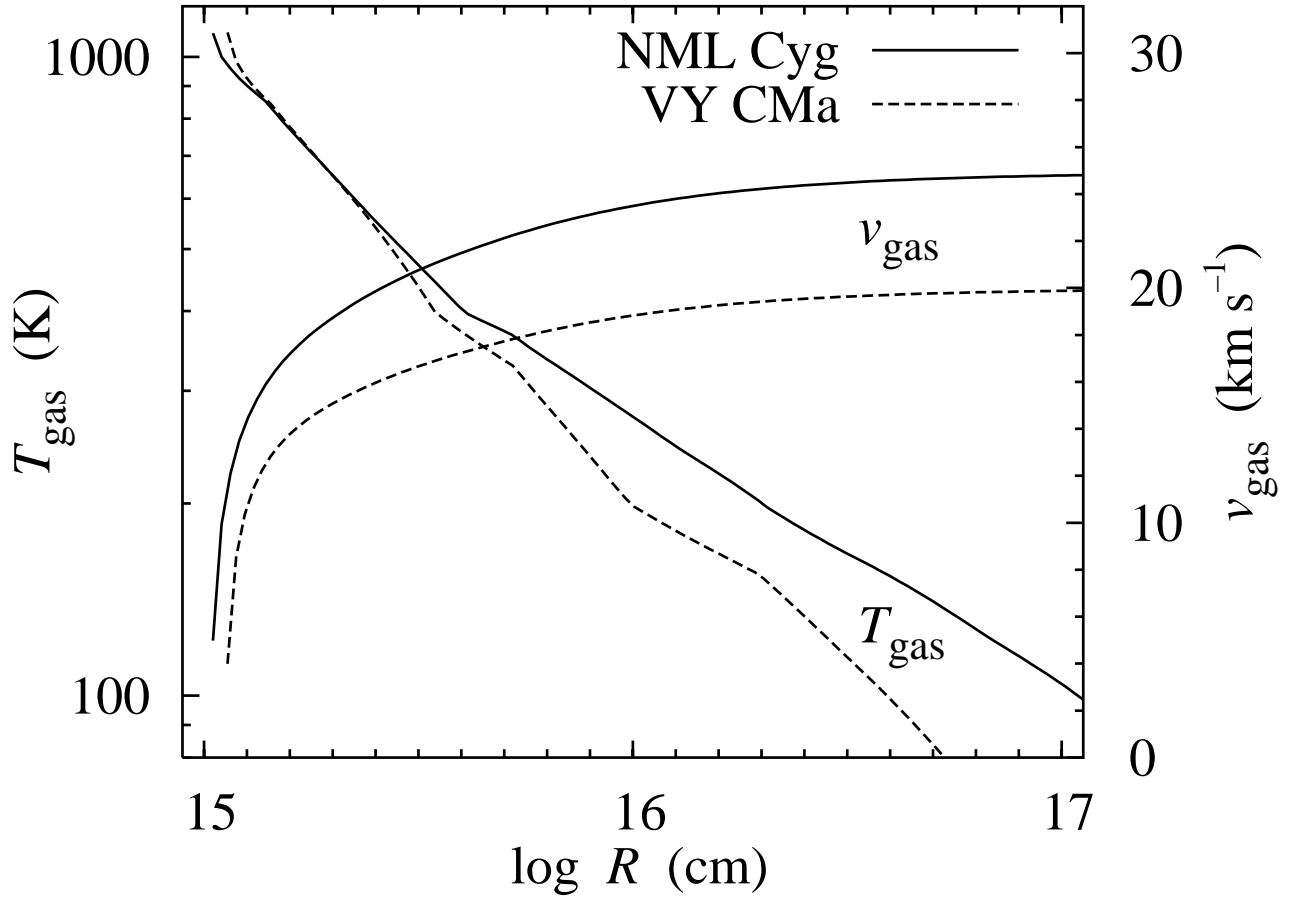}}
\caption{Velocity and temperature profile of the gas flowing out of NML Cyg and VY CMa,
derived with the
model of Zubko \& Elitzur (2000).}
\label{fig:Outflow}
\end{figure}
\end{document}